\documentclass{article}

\usepackage{graphicx}
\usepackage{amsthm,xcolor}
\usepackage{amsfonts}
\usepackage{amsmath}
\usepackage{amscd}
\usepackage{amssymb}
\usepackage{alltt}
\usepackage{url}
\usepackage{ellipsis}
\usepackage{lineno}
\usepackage{setspace}
\usepackage[numbers]{natbib}

\theoremstyle{thmstyleone}%
\theoremstyle{thmstyletwo}%

\theoremstyle{thmstylethree}%

\begin{document}

\title{Robert Millikan, Japanese Internment, and Eugenics}

\author{Thomas Hales} 
\date{}

\maketitle

\abstract{Robert A. Millikan (1868--1953) was the second American to win the Nobel
Prize in physics.  At the peak of his influence, no scientist save
  Einstein was more admired by the American public.

Millikan, the head of the California Institute of Technology (Caltech)
during its first 24 years, oversaw its rapid growth into one of the
leading scientific institutions of the world.  However, in response to demands
for social justice following the murder of George Floyd, Caltech
launched an investigation into Millikan.  Caltech reached a decision
to strip Millikan of honors (such as the library named after him),
following accusations from various sources that he was a sexist,
racist, xenophobic, antisemitic, pro-eugenic Nazi sympathizer. In
short, Caltech threw the book at him.

This article analyzes two accusations against Millikan.
The first of these accusations
was published in \emph{Nature}: that he collaborated to deprive Japanese
Americans of their rights during their forced relocation to internment
camps during the Second World War.
An examination of original historical sources will show that this
accusation is false.  On the contrary, Millikan actively campaigned
during the war to promote the rights of Japanese Americans.
This article traces the stages of misrepresentation that led to
current false beliefs about Millikan.  In view of Millikan's
extraordinary position in American science, this misrepresentation is
a cautionary tale.

The article also treats Caltech's central accusation against
Millikan: he lent his name to ``a morally reprehensible
eugenics movement'' that had been scientifically discredited in his
time.  The article considers the statements purporting to show that
eugenics movement had been denounced by the scientific community by 1938.  
In a reversal of Caltech's claims, all three of Caltech's scientific
witnesses against eugenics -- including two Nobel laureates -- 
were actually pro-eugenic to varying
degrees.  This article concludes that Millikan's beliefs fell within 
acceptable scientific norms of his day.

}

\maketitle


\section{Robert Millikan}


\subsection{Scientific contributions}

Millikan's life spanned one of the most fertile periods in the history
of physics, covering the birth of quantum theory, special and general
relativity, subatomic physics, and nuclear energy.  He was born in a
small town in Illinois in 1868 not long after James Maxwell published
his famous equations on electromagnetism and died in 1953 shortly
before the birth of Yang-Mills theory.  Millikan was one of the
greatest experimental physicists in the world during the first half of
the twentieth century.  He built his reputation on high-precision
instruments that he designed for experimental use.  During his
lifetime, the United States moved from the margins of physics to a
powerful central position.

In 1951, in a letter to Werner Heisenberg, Millikan made a
self-assessment of what he thought to be his most important scientific
contributions~\cite[43:729]{Millikan}.  First on his list was the
isolation the electron and the measurement of its charge.  Millikan
was awarded a Nobel Prize in 1923 for the famous oil-drop experiment,
which measured the electron charge.  These measurements also led to
reliable estimates of Avogadro's number~\cite{13936}.

Second was his experimental verification of Albert Einstein's
photoelectric equation, which was also recognized by the Nobel
Prize.  Einstein's and Millikan's Nobel Prizes are closely linked:
Einstein's for the theoretical derivation the photoelectric equation
and Millikan's for the experimental verification.  (Contrary to what
some might expect, Einstein's Nobel Prize was not for the theory of
relativity.)  As a bonus, Millikan obtained what was then the best
experimental value of Planck's constant.

The photoelectric effect gave one of the first hints of wave-particle
duality.  Einstein hypothesized the existence of what is now called a
photon -- a hypothesis that flew in the face of the settled science of
wave optics.  This unsettling concept of light motivated Millikan in
his experiments, and he eventually confirmed Einstein's equation.
At the celebration of
Einstein's seventieth birthday in 1949, Millikan recalled
\cite[68:642]{Millikan}~\cite[p.~106]{millikan2020autobiography}:
\begin{quote}
I spent ten years of my life testing that 1905 equation of Einstein's,
and, contrary to all my expectations I was compelled in 1915 to assert
its unambiguous experimental verification in spite of its
unreasonableness since it seemed to violate everything that we knew
about the interference of light.  The contradictions involved in this
equation could not be removed by any considerations which were
available at the time to Planck, to Einstein or to any of the rest of
us. These contradictions have now largely disappeared, however,
through the development of the so-called ``wave mechanics'' by the
work of Louis de Broglie, Schr\"odinger, Heisenberg, and Dirac.%
\footnote{
According to science historian Allan Franklin,
Millikan's interpretation of his own experiment evolved over time.
At the time, ``neither
Millikan, nor the physics community accepted'' Millikan's experiments
as a confirmation of the photon theory~\cite{Franklin}.
The quotation expresses Millikan's late view.}
\end{quote}
Even today, wave-particle duality continues to pose some puzzling
philosophical questions~\cite{maudlin}.  Millikan's research on the
photoelectric effect and the electron helped to propel physics from
the atomic scale to subatomic particles.


Third was the closing of the gap between light spectra and X-ray
spectra, using the ``hot spark source of light.''  
Robert Kargon described
the research findings  as follows:
\begin{quote}
In a series of papers with R.A. Sawyer and Ira Bowen, Millikan was
able to make a considerable extension of the map of the ultraviolet
spectrum; they had been able to photograph, measure the wavelength,
and analyze the atoms of light elements and multiply ionized atoms of
the heavier atoms. They found about 1,000 new [atomic spectral] lines,
and showed that their wavelengths were consistent with the Bohr
theory --Kargon~\cite[p.~125]{kargon2020rise}.
\end{quote}
Millikan and his collaborators compared their new
lines against the predictions of an enhancement of the Bohr model: the
relativistic Bohr-Sommerfeld model of the atom~\cite[pp.
  209--231]{millikan1947electrons}.  They found some problems with the
theory that were later reconciled by the introduction of electron spin
to the model~\cite{uhlenbeck1926spinning}.%
\footnote{The Uhlenbeck-Goudsmit article of December 1925, which
introduced the concept of electron spin, stated, ``The assumption of
the spinning electron leads to a new insight into the remarkable
analogy between the multiplet structure of the optical spectra
and the structure of X-ray spectra, which was emphasized especially
by Land\'e and Millikan.\,\ldots\ [This analogy] obtains an
immediate explanation on the hypothesis of the spin electron''~\cite{uhlenbeck1926spinning}.}

The fourth on the list is ``the law governing the pulling of electrons
out of metals by intense electrical fields.''
See~\cite[p.~126]{kargon2020rise}
and~\cite[p.~261]{millikan1947electrons}. The experimental results of
C. F. Eyring, Millikan, and Charles Lauritsen were eventually explained by
J. Robert Oppenheimer quantum-mechanically as electrons tunneling through a
potential barrier.

Fifth was the extraterrestrial origin of cosmic rays.  The term
\emph{cosmic rays} was coined by Millikan, but he was not the first to
provide evidence of their extraterrestrial origin.  In
1912, several years before Millikan became involved in cosmic ray research,
Victor Hess, in Nobel Prize winning work, made measurements in
balloons and ``concluded that the upper atmosphere is ionized by
radiation from space''~\cite{hess}.  Millikan and G. Harvey Cameron
submerged electroscopes into mountain lakes and found that their
readings at different water depths confirmed the extraterrestrial
origin of cosmic rays~\cite{dumond}.  In the early days of cosmic ray
research there was much to investigate: how cosmic ray intensity
varies according to altitude, latitude, and longitude; how deeply
they penetrate; whether the rays carry positive, negative, or no charge;
and how the energy is distributed.  Millikan was among the most active
researchers in this field.

Millikan was sometimes wrong. If cosmic rays carry charge, then they
are deflected by the earth's magnetic field, producing a more abundant
shower of cosmic rays near the earth's poles than the equator.
Millikan and Cameron made a scientific expedition to Peru in 1926 to
answer this question experimentally, but because of their lack of
thoroughness, limitations of their electroscopes, and partial
equipment failure, they failed to detect any significant dependence of
cosmic ray intensity on latitude.  Millikan fatefully took
non-detection as evidence that cosmic rays carry no charge.  In
opposition to Millikan's conclusion, the next year, Jacob Clay
presented experimental evidence that cosmic rays do indeed vary in
intensity according to latitude, and hence that cosmic rays do carry
charge.  Arthur Compton backed up Clay's results.  Millikan eventually
accepted the Clay-Compton conclusion, but only slowly and only after
acrimonious exchanges with Compton, which were dragged into the public
arena by \emph{New York Times} reporting~\cite{kargon2020rise}.

Sixth is the design of a specialized cloud-chamber for the detection of cosmic
particles.  Carl Anderson was a PhD student under Millikan at Caltech.
After completing his PhD, Anderson continued at Caltech as a
Research Fellow (1930--1933) but shifted to cosmic ray research, still
supervised by Millikan, who ran three different cosmic ray
research groups, each using a different kind of detector.  
Anderson later recalled, ``Professor R. A. Millikan and the writer in
the spring of 1930 planned a cloud-chamber apparatus suitable for
cosmic-ray studies,\,\ldots\cite{anderson}''%
\footnote{Millikan wrote that he and Anderson designed the cloud chamber
in the summer of 1929, but Anderson's date seems more plausible,
because of the period of Anderson's
fellowship~\cite[p.~322]{millikan1947electrons}.}  Anderson built the
cloud chamber over a period of months. In August 1932, when Anderson
was only twenty-six years old, his cloud chamber detected a positron, the
first form of anti-matter ever detected.  For the discovery of the
positron, Anderson was awarded a Nobel Prize in physics. It was
Millikan who nominated him.  Four years later, Anderson and his first
graduate student, Seth Neddermeyer, discovered yet another elementary
particle, the muon, which carries the same electric charge as the
electron, but possesses greater mass.

\subsection{Secondary influences}

Millikan attracted and was attracted to truly extraordinary scientific
talent.  He went to Germany for three semesters after finishing his
PhD at Columbia in 1895 and was present in Berlin for Wilhelm
R\"ontgen's first large public exhibition of the X-ray.  In 1896,
Millikan attended Max Planck's lectures. According to Millikan, it was
during those lectures that Planck conceived the idea of the quantum or
discontinuous change (but did not publish on the topic until a few
years later).  Elsewhere in Europe were Henri Becquerel's discovery of
radioactivity in 1896 and J. J. Thomson's research on the electron
in 1897.  It was a new era in physics. Millikan wrote in his
autobiography that these ``discoveries actually determined the
direction of my own study and research for the next fifty
years''~\cite[p.~270]{millikan2020autobiography}.


The first American Nobel Prize winner in physics was Albert
A. Michelson, who is remembered for the famous Michelson-Morley
experiment, which failed to detect motion of the earth relative to the
luminiferous ether, a fact later explained by Einstein's special
theory of relativity.%
\footnote{Michelson's Nobel Prize was awarded for ``for his optical
  precision instruments and \ldots\ investigations carried out with
  their aid.''  }  
It was Michelson who recruited Millikan to the
University of Chicago.  For decades in Chicago, until Millikan left
for Caltech, the two were close colleagues and friends, playing tennis
together regularly~\cite[p.~87]{millikan2020autobiography}.  Michelson
regarded Millikan as his successor at the University of Chicago, but
history had other plans.

The second American to win a Nobel Prize in physics was Millikan.  The
third was Arthur Compton, Millikan's nemesis in the cosmic ray debate.
(Arthur's brother Karl will be relevant later in this article.)  The
fourth was Millikan's student Carl Anderson, mentioned earlier. The
fifth was Clinton Davisson, who as a student at the University of
Chicago was inspired by Millikan to go into physics.  Davisson faced
financial hardship as a student, and it was through Millikan's
recommendation that Davisson obtained employment as a physics teacher,
while he continued to work his way through
school~\cite{gehrenbeck1978electron}~\cite{kelly1962clinton}.
Davisson won his Nobel Prize for observing diffraction patterns of
electrons, confirming Louis de Broglie's prediction that wave-particle
duality is not limited to photons.


Millikan's influence extended beyond pure science into
industrial research that shaped the modern world.  The physicist Frank
Jewett was the first president of the legendary AT\&T Bell Telephone
Labs starting in 1925 and later became the chairman of its board of
directors.  Jewett and Millikan were the closest of friends.  Millikan
helped line up Jewett with his wife, who had been a Millikan student; and
Jewett was the best man at Millikan's wedding~\cite[p.
53]{millikan2020autobiography}.


In 1909, chief engineers and executives at AT\&T started discussing
the feasibility of establishing transcontinental phone service.  The
core technological problem was that as the telephone signal travels
down the copper wire it grows weaker, so that a signal originating in
New York would be entirely lost before reaching San Francisco.  What
was needed was an efficient \emph{amplifier} (or \emph{repeater} to
use the term originating in the telegraph industry) that boosted the
signal and kept it from attenuating in its transcontinental
journey~\cite[p.~21]{gertner2012idea}.

Already in 1909, Jewett was a senior manager at AT\&T.  The task of
amplification became Jewett's responsibility.  He consulted his friend
Millikan about the problem in 1910.  Jewett asked Millikan to let him
``have one or two, or even three of the best young men who are taking
their doctorates with you and are intimately familiar with your field.
Let us take them into our laboratory in New York and assign to them
the sole task of developing a telephone repeater''~\cite[p.
22]{gertner2012idea}.  Millikan sent Jewett his recent PhD student,
Harold Arnold (who would later become Bell Lab's first director of
research) to work on the task.


What happened next is mythical.  An amplifier, called an
\emph{audion}, had already been recently invented by Lee de Forest.
(In 1946, Millikan called de Forest's audion the single most important
advance in electronics of all time~\cite[43:63]{Millikan}.)  Arnold
and his team went to work improving the performance of the audion for
AT\&T, after the organization bought the patent rights to de Forest's
invention.  From time to time, Millikan consulted on the project, but
he was mostly a bystander.  The research and development resulted in a
practical vacuum tube amplifier, which was used in the
transcontinental telephone line.  At the Panama-Pacific International
Exposition in 1914, Alexander Graham Bell and Thomas Watson,
positioned in New York and San Francisco, gave the first
transcontinental public demonstration: ``Mr. Watson, come here, I want
you,'' echoing their immortal words from decades
earlier~\cite[pp.~23--24]{gertner2012idea}.  The vacuum tube amplifier
revolutionized electronics throughout radio, film, television,
computer, and consumer electronics.


The friendship between Millikan and Jewett led to a procession of
physicists that were trained by Millikan and employed by Bell Labs.
Mervin Kelly was another physicist who obtained his PhD under
Millikan~\cite[p.~16]{gertner2012idea}.%
\footnote{How many PhD students did Millikan
supervise?  A very rough estimate suggests he supervised more than
eighty PhDs in physics, including more than thirty from the University of Chicago.  At
Caltech alone, by 1940, Millikan supervised about one-third of the
more than 135 PhDs~\cite[p.~107]{goodstein2020millikan}.}  Kelly became
director of research at Bell Labs starting in 1936 and later became
president of Bell Labs.  As research director, Kelly redirected
research focus from vacuum tube technology to solid-state physics and
recruited William Shockley to lead the solid-state research group.
Kelly once stopped by Shockley's office (which was shared with
Davisson) and lectured on the coming day when something electronic
would replace telephone relays. ``For the rest of his life Shockley
considered Kelly's lecture as the moment when a particular idea freed
his ambition, and in many respects all modern technology from its
moorings''~\cite[p.~23]{gertner2012idea}.  The research group --
Schockley, John Bardeen, and Walter Brattain -- was awarded the Nobel
Prize in physics for the invention of the transistor.

Relationships were not merely scientific.  Chien-Shiung Wu is known
for the discovery of parity violation, for which she was awarded the
Wolf Prize in physics in 1978. When Chien-Shiung Wu married Luke
Chia-Liu Yuan in 1942, none of the parents were able to attend,
because of the war.  The Millikans hosted the wedding, which was held
at their home.  Millikan was Yuan's doctoral advisor.  Yuan's
grandfather was Yuan Shikai, the first president of the republic of
China.

\subsection{Education}

In education, Millikan and Gale's \emph{First Course in Physics} is
perhaps the best selling English-language physics textbook of all
time.  Including all editions and title variations, Millikan and Gale
sold 1,610,637 copies between 1906 and 1952~\cite[43:520]{Millikan}.
It is particularly remarkable that these numbers were achieved in the
first half of the twentieth century, when the physics textbook market
was much smaller than it is now.  In the international textbook
market, a Caltech webpage states that ``\emph{The Feynman Lectures on
Physics} is perhaps the most popular physics book ever written.  More
that 1.5 million English-language copies have been sold; probably even
more copies have been sold in a dozen foreign-language editions (the
number of copies in Russian alone, for example, is estimated to be
over 1 million)''~\cite{feynman}.  By way of comparison, popular
physics books have sold far more copies.  Stephen Hawking's
\emph{A Brief History of Time} has sold more than 25 million copies~\cite{mckie2007brief}.


\subsection{Caltech's executive}

Kargon wrote that the physicist Millikan, astrophysicist George Hale,
and chemist Arthur Noyes formed a ``triumvirate'' that ``was
responsible for the rapid rise to prominence of\ \ldots\ the California
Institute of Technology''~\cite{kargon2020rise}.  
The idea of a founding triumvirate of Caltech has been
repeated almost to the point of clich\'e, but the three scientists
contributed in entirely different ways to the rise of the institute.


George Hale was the dreamer. He called himself a schemer.  In 1903,
Hale secured funding from the Carnegie Institute to construct a solar
observatory on Mount Wilson near Pasadena.  Each of his dreams led to
another.  Hale envisioned a chemical laboratory to handle problems
that arose at Mount Wilson, then more ambitiously of an outstanding technical
institute to house the chemistry lab.  In 1907, Hale became a trustee
of Throop Polytechnic Institute in Pasadena, which was an unremarkable
local manual training school. Hale channeled his hope of creating an
outstanding institute into Throop. He spent years recruiting Noyes to head
the division of chemistry, and Millikan to serve as president.  Jewett
called Hale the ``gifted strategist'' and Millikan ``the field
general''~\cite[p.~92]{kargon2020rise}.  What Hale schemed, Millikan
brought to life.


Millikan was appointed the chief executive of Caltech in 1921, the
year after Throop was renamed the California Institute of Technology;
he held the position for twenty-four years. He was offered the title of
president but instead chose an organizational structure that made
him the chairman of an executive council, consisting of four trustees
and four distinguished members of the faculty. Noyes was one of them.
Millikan was also the director of Caltech's physics division -- the Norman
Bridge Laboratory of Physics.  ``Millikan was everywhere planning,
pushing, deciding, admonishing.\,\ldots\ His day at the Institute,
starting at eight o'clock in the morning, frequently did not terminate
until long after midnight''~\cite{dumond}.

Noyes had been acting president of MIT for two years before moving to
Caltech and had considerable administrative know-how.  Linus Pauling
wrote that ``Millikan became a great public figure, who in the minds
of the people of the country represented the California Institute of
Technology; but Noyes was often the one who was responsible for the
policies that were announced by Millikan''~\cite{pauling1958arthur}.


When Millikan first arrived at Caltech, DuMond recalled, ``The faculty
and graduate students were still a small enough group so that at the
first faculty dinner we could all sit around a single long table in
the basement of the Crown Hotel''~\cite{dumond}.
Under Millikan, Caltech swiftly grew.
During its first decade, Caltech made several stellar faculty appointments.
\begin{itemize}
\item Theodor von Karman, who became the director of Caltech's
Aeronautical Laboratory and later a founder of the  Jet Propulsion
Laboratory;
\item Carl Anderson, the Nobel Prize winning discoverer of the positron, 
discussed earlier; 
\item Fritz Zwicky, an astronomer of
neutron stars, dark matter, and gravitational lenses;
\item Robert Oppenheimer, who was head of the Los Alamos Lab during
the Manhattan Project to develop nuclear weapons;
\item Richard Tolman, who served as scientific advisor to General
Leslie Groves, the director of the Manhattan Project;
\item Thomas Hunt Morgan, who
won the Nobel Prize in 1933 for establishing that chromosomes carry
the genetic material;
\item Alfred Sturtevant, who made the first genetic map of a chromosome;
\item Theodosius Dobzhansky, who was one of the architects of the
modern synthesis, combining Darwinian evolution, population
genetics, and Medelian genetics; and
\item Linus Pauling, one of the founders of quantum chemistry 
and the recipient of two Nobel Prizes (in Chemistry and Peace).
\end{itemize}
Other renowned scientists were closely affiliated with Caltech.  Charles
Richter, who developed the Richter scale, was part of the Caltech
Seismological Lab, which at the time was a cooperative venture between
the Carnegie Institute and Caltech.  Edwin Hubble, who gave evidence
for the expansion of the universe, was nearby at the Mount Wilson
Observatory.

The growth of Caltech can be measured in many ways.  In
scientific productivity, ``by 1930, Caltech was ranked as the leading
producer of important physics papers in the
country''~\cite[p.~108]{goodstein2020millikan}.  Its endowment, which
was almost nonexistent in 1920, had grown to \$25 million by 1947.
Caltech had completed construction of only two permanent buildings in
1920 and thirty-six by 1947~\cite[p.~249]{millikan2020autobiography}.


Caltech initiated a broad array of scientific projects, each of which
was an undertaking of industrial scale, requiring major funding (from
sources such as the Rockefeller Foundation, the Guggenheim Fund, the
Carnegie Institution, or industrial partners); scientists, engineers,
and students; buildings, labs, and equipment.  In cooperation with the
Southern California Electric Company, a high-voltage lab was built.
The Guggenheim aeronautics lab tested Douglas airplanes in its wind
tunnel.  ``Caltech's Daniel Guggenheim Graduate School of Aeronautics
played a major role in turning southern California into the aircraft
capital of the world~\cite[p.~177]{goodstein2020millikan}.''  There
were Millikan's cosmic ray project, a seismology project, the study
and development of jet propulsion, and the study of oil under high
pressure sponsored by the American Petroleum Institute.  Starting in
the 1920s, Hale had a vision of building the largest telescope in the
world.  Caltech's Palomar Observatory, including its $200$-inch Hale
Telescope, became operational in 1948~\cite[pp.
238--250]{millikan2020autobiography}.

\section{Japanese Internment and the Fair Play Committee}

World-altering political events shaped Millikan's final years at the
head of Caltech.  After the attack on Pearl Harbor, the United
States declared war on Japan.  Japanese Americans living on the West
Coast were forced to relocate to internment camps.  An organization
named the \emph{Fair Play Committee} was formed in Berkeley to defend
the rights of Japanese Americans.  Millikan became a vice president in
this organization.  After the war ended, scientific missions were
organized to Japan to evaluate Japan's scientific capabilities.
Millikan assisted in the recruitment of scientists and engineers for
these missions.

\subsection{Pearl Harbor}

Japan attacked Pearl Harbor on December 7, 1941.  The next day, the
United States declared war on Japan.  In January 1942, anti-Japanese
sentiments intensified, and widespread hostile feelings were directed
against Japanese Americans.%
\footnote{This article uses the term \emph{Japanese American} inclusively both for
United States immigrants of Japanese citizenship (Issei), and for
American-born citizens (Nisei) of Japanese ancestry.  What this
article and many sources today call internment camps were formerly
called relocation projects. Formerly, internment camps meant enemy
alien camps administered by the Department of Justice.}
Anti-Japanese rhetoric was especially fierce in California, which was
the home to the majority of all continental Japanese Americans.
In California, the Hearst newspapers, the McClatchy newspapers, the 
\emph{Los Angeles Times}, as well as hundreds of civic organizations
were united against Japanese Americans~\cite[p.
37]{reeves2015infamy}~\cite{kingman}.  On January 29, 1942, a Hearst
newspaper columnist wrote of the Japanese on the West Coast, ``Herd
'em up, pack 'em off and give 'em the inside room in the badlands.
Let 'em be pinched, hurt, hungry and dead up against it.\,\ldots\
Personally, I hate the Japanese. And that goes for all of
them''~\cite{herd-em}.  A barber shop offered ``free shaves for Japs''
but was ``not responsible for accidents''~\cite[p.
18]{thomas1974spoilage}.

The newspapers and public opinion were backed up by political power.
In February 1942, Earl Warren was the California attorney general and
was soon to be elected governor of California (January 1943--1953).
Years later, he would become the chief justice of the United States.
Warren favored the evacuation of Japanese aliens from California.

Walter Lippmann,
one of the most respected journalists of his time, visited
Warren and General John DeWitt in California. Lippmann published an
enormously influential column that appeared in the \emph{Washington
Post} and 250 other newspapers on February 12, 1942.%
\footnote{After meeting in person with him, Millikan praised
  Lippmann's 1943 book on U.S. foreign policy for its clarity of
  thought~\cite[45:889]{Millikan}. However, in 1942 on
  Japanese-American policy their views were in sharp opposition.  }
The article amplified the general's and attorney general's fears of
sabotage.  The title ``The Fifth Column'' referred to those who engage
in sabotage within their own country.  Lippmann stoked fears of a
Japanese sea and air attack on California, coordinated with sabotage
by ``enemy agents on land.'' His readers knew which racial group he
was referring to.  He called for Washington to act decisively to
establish the West Coast as a combat zone, letting ``the explanations
and the reparations come
later''~\cite{lippmann}~\cite[pp.~49--50]{reeves2015infamy}.
Lippmann's column went to Secretary of War Henry Stimson and to the
second in command, John J. McCloy.  According to McCloy's biographer
Kai Bird, ``More than any other individual, McCloy was responsible for
the decision [of Roosevelt], since the president had delegated the
matter to him through Stimson''~\cite{bird2017chairman}.

On February 19, 1942, a little more than two months after the attack
on Pearl Harbor, President Roosevelt signed Executive Order 9066, which
authorized the establishment of military zones;  all
Japanese Americans living within these zones were soon to be forcibly
expelled and relocated to internment camps.  The military zone
included all of California, coastal portions of Oregon and Washington,
and southern parts of Arizona.


Starting in 1942, all  Japanese Americans (about 112,000) were forcibly
relocated from the West Coast into ten internment camps that were
built in scattered locations throughout western states and Arkansas.
For example, the Topaz internment camp in central Utah housed over
8000 Japanese Americans in military-style barracks, each small family
in a one-room apartment lit by a single lightbulb and heated by a
potbelly stove.  Life was austere, yet the camp offered a range of
basic services: a hospital, library, schools, Buddhist and Christian
churches, general stores, co-ops, banks, repair shops, and police and
fire departments run by the internees.  The camp was one square mile,
surrounded by barbed wire and guard stations.


A civilian government agency, the War Relocation Authority (WRA), ran the
internment camps. Dillon S. Myer was the WRA director for almost the
entire duration of agency's existence, 1942--1946.  Although the director
over internment, Myer became opposed to internment.%
\footnote{Assessments of Myer vary enormously.  Greg
Robinson's book on internment states that Japanese Americans
celebrated Myer's camp administration during the war, but that
Japanese-American attitudes turned sharply against him in the
1980s~\cite[p.~256]{robinson2001order}.
Others held that Myer inherited ``what may well have been the
toughest, most exasperating civilian job during the
war''~\cite{densho-myer}~\cite{weglyn1976years}.  }  
The historian
Alonzo Hamby, in a book review of Greg Robinson's book on
internment \cite{robinson2001order}, wrote that Myer and a few others
``struggled futilely for an early end to the program against a flood
tide of hysteria and expediency''~\cite{hamby}.

Over time, legal challenges against internment made
their way to the Supreme Court.  Finally, on December 17, 1944, the
``Roosevelt administration issued Public Proclamation No. 21\ \ldots\ 
declaring that Japanese Americans could return to the West Coast the
next month.''  This proclamation marked the end of Japanese
internment.  The proclamation pre-empted by one day the Supreme Court,
which issued two decisions (Korematsu
v. United States and Ex Parte Endo). The justices ruled that
the WRA had no authority to detain loyal citizens.


\subsection{Japanese-American petitions to Millikan}

When Japan attacked Pearl Harbor on December 7, 1941, Robert Millikan
and his wife Greta were traveling in Mexico City.  As Greta tells it,
``A Mexican gentleman, recognizing us as Americans, I suppose, told
[our guide] E.  that he had just heard on the radio of Japan's attack
on Manila and Hawaii.  It seems impossible that it has really come --
we were sobered and stunned,\,\ldots''~\cite[80:89]{Millikan}.

As soon as Greta was back in Pasadena in January 1942, she hurried
over to the home of her Japanese gardener Harris Ozawa and his family,
recording in her diary, ``Although warned to stay at home, they had
gone quietly about their jobs, having nothing to conceal or be ashamed
of''~\cite[80:108]{Millikan}.  Later, when about to be forced out of
Pasadena, in a sentimental scene, Ozawa replanted the azaleas from his
own home into the Millikan garden as a parting gift, but Greta
insisted that the flowers were only on loan until his return.  Greta
saw the Ozawas off at the train station, as they left for the
Tulare detention camp: ``With duffle bags and big suit cases, bedding
rolls and baskets, families assembled at the starting point where a
sympathetic social service worker helped them with details; a group of
church women served hot coffee and rolls.''  Concerned about the
living conditions at the temporary detention camps, Greta visited the
Santa Anita camp.  She made contacts with organizations to aid
Japanese-Americans.  Harris's wife Elizabeth wrote frequent letters to
Greta that chronicled life under internment.  After the war, the Ozawa
family was prevented from retaking their house in Pasadena, and Greta
assisted with the ``time-consuming and not a little exasperating''
legal process to regain possession~\cite[80:109,132,135,284,381]{Millikan}.

The war brought Robert Millikan into renewed contact with Japanese
acquaintances.  A friend, T. Hori, sent a plea to
Millikan~\cite[43:1016]{Millikan}.

\begin{quote} 
Yes, the United States is now at war.
It is my sincere hope that this country will win and forever
banish the demons of war from this world.\,\ldots

Last Monday, a week ago today, all our assets were frozen
because of my visit to Japan for the sole purpose of visiting
my invalid brother who lived in this country for over thirty
years, and also because I am an alien.  During this past
week, a majority of the leading Japanese aliens were taken
away to the Terminal islands for investigation,\,\ldots

On the other hand, my wife, who is an American citizen,
received many telephone calls from her American church
friends asking about her welfare.\,\ldots\ This kindness and
thoughtfulness impresses us deeply.\,\ldots

Yes my entire household is wholeheartedly supporting the
cause of freedom and democracy for which the United States
stands.\,\ldots\quad -- Hori, Dec 15, 1941
\end{quote}

Arriving back in Pasadena from Mexico, Millikan responded to
Mr. Hori~\cite[43:1016]{Millikan}:

\begin{quote}

Your letter touched me deeply.  I shall be glad to talk over
the situation with you anytime, and if I can be of any assistance
in making the difficult situation in which you and other loyal
members of your group find yourselves I shall want to do it.
Mrs. Millikan, too, has been wondering whether she could be of
help to members of your group.
-- Millikan, Jan 6, 1942
\end{quote}

A former Caltech student Koichi Kido sent a plea 
to Millikan on January 28~\cite[28:452]{Millikan}:

\begin{quote}
You, Dr. Millikan, are a champion of democracy. You, sir,
are a champion of human freedom.  Will you fight to help
maintain them in these United States?\,\ldots

There is the probability that many of us will be moved
inland very shortly.\,\ldots There is the final possibility
that all of us will not only be moved inland but be interned
for the duration.\,\ldots

We who are citizens of the United States just as you are and
who have no political ties with Japan (dual citizenship for
instance) have the same rights and privileges as you do.  
Everyone who has any knowledge of the Constitution and the
Bill of Rights is aware of this fact.\,\ldots

Who will defend us?
-- Koichi Kido, Jan 28, 1942
\end{quote}

Moved by Koichi Kido's letter,
Millikan wrote to the Pasadena Chamber of Commerce and
City Board of Directors~\cite[28:454]{Millikan}.

\begin{quote}
Pearl Harbor awakened this country from its idle dream
of isolated security.  It is very important that we do
not go to the other extreme and take hysterical, instead
of considered, action in the matter of defense.\,\ldots

It is particularly important that we do not go to any unwise wholesale
schemes in trying to protect ourselves against the dangers arising
from our residents of enemy country origin.\,\ldots

To adopt any wholesale policy of putting them all in concentration
camps would be an action which would defeat its own purpose, not only
in weakening the defense industries in which they are so useful, but
also in arousing justified resentment because of the unfair treatment
of loyal American citizens and thus increasing, rather than decreasing
the danger of sabotage.  -- Millikan, Feb 4, 1942
\end{quote}

Millikan responded to Mr. Kido~\cite[28:454]{Millikan}.

\begin{quote}
[I] appreciate fully the difficult situation in which you,
the American citizens of Japanese parentage, are placed by
the action of the Japanese government.  I have been doing all
I can to get discrimination and sanity into the treatment of the many
who are likely to suffer unjustly in this difficult situation.
If there is anything in particular I can do to help you
I hope you will let me know.
-- Millikan Feb 6, 1942
\end{quote}

By mid-February, through two of its press releases, Millikan had
become aware of the Fair Play Committee in the California Bay Area, which
opposed discrimination against Japanese civilians.  (The organization
was to go through several renamings, but this article consistently
calls it the Fair Play Committee.)  The organization's first press release in October
1941 (before Pearl Harbor) called upon ``fair-minded Californians to
combat discrimination against their fellow residents of Japanese
race''~\cite{kingman-brief}.  The first paragraph of the press release
stated~\cite{kingman-brief}.

\begin{quote}
\ldots\ popular
resentment toward Japan may find expression in greater discrimination
or even physical violence against fellow-residents of Japanese
extraction, distrust of the Japanese Government being transferred
to all persons of Japanese race.  A moment's thought will show
that such animus would be not only un-American, but also a 
menace to public welfare and the good name of the state.
-- David Barrows, Fair Play, Oct 1, 1941.
\end{quote}

On February 15, 1942,  Millikan wrote to the Fair Play organization
with an interest ``in securing cooperation in Los Angeles County.''
Millikan's overture was welcomed.
On March 3, 1942, Millikan was appointed a vice president
of the committee.  Millikan was now a senior officer
in an organization fighting for the fair treatment of Japanese
Americans.

\subsection{Fair Play}

The Fair Play Committee was formed in the fall of 1941 by David Barrows
and Galen Fisher to defend the rights of Japanese Americans.
Barrows was a political scientist and
former president of the University of California.  
According to the historian Robert Shaffer~\cite{densho-fisher},
\begin{quote}
Galen Merriam Fisher (1873--1955) was probably the most significant
and consistent white organizer of opposition during World War II to
the wholesale incarceration\ \ldots\ of Japanese Americans.  As a
former missionary in Japan\ \ldots\ Fisher was the key founder in 1941
of the [Fair Play Committee].\,\ldots\ Throughout the war Fisher wrote
numerous articles criticizing the mistreatment of Japanese
Americans. -- Robert Shaffer, Densho Encyclopedia
\end{quote}
From the beginning, the committee had strong university ties,
especially with Berkeley.  Robert Sproul, the president of the
University of California, became the honorary chairman.  Provost
Monroe Deutsch and several Berkeley professors were also active
participants.  


The Fair Play advisory board consisted of a group of highly
influential political, business, educational, and religious leaders,
including Stanford's Chancellor Ray Lyman Wilbur;
its Dean of the Graduate School of Business, Jackson; former
California Governor, C. C. Young; the President of the California
State Chamber of Commerce, A. Lundburg; former President of the State
Bar Association, G. Hagar; the Mayor of Berkeley, F. Gaines; and many
others.

The Committee's executive secretary, Ruth Kingman, had
access to the corridors of government power, meeting with key
government figures such as California State Attorney General Bob
Kenny, U.S. Attorney General Biddle, WRA Director Myer, Assistant
Secretary of War McCloy, FDR's daughter Anna Roosevelt Boettiger, and
some members of Congress.%
\footnote{Robert Cozzens, the Assistant National Director of the WRA, said that
``Ruth Kingman was to see us at least once a week and generally a
couple of times a week.''  Kingman worked a lot with
state Attorney General Bob Kenny, who was ``said to have influenced
Warren finally to accept the inevitable''~\cite{cozzens}.}
Judicious and steadfast, she always seemed to know what to suggest
next to nudge policy makers one step further into alignment with the
goals of Fair Play.  The committee was loyal to the United States and
the war effort and worked as a ``moderating influence on both public
opinion and government authorities, and helped avert mob violence
against Japanese residents''~\cite{kingman-brief}.  The committee was not a relief
organization. It worked to change policy and public opinion through
speakers, educational materials, and corrections to the press.
A membership drive leaflet affirmed its purpose:
\begin{quote}
The fundamental purpose of the Committee is to support the
principles enunciated in the Constitution of the United States,
and to that end to maintain, unimpaired, the liberties guaranteed in
the Bill of Rights.  The Committee believes:
\begin{enumerate}
\item That attacks upon the rights of any minority tend to undermine
the rights of the majority.
\item That attempts to deprive any law-abiding citizen of his citizenship
because of racial descent are contrary to fundamental American principles
and jeopardize the citizenship of others.\,\ldots
\item That it is un-American to penalize persons of Japanese descent
in the United States solely for the crimes of the Government 
and military cast of Japan.
\end{enumerate}
-- Fair Play~\cite{fair-play-drive}
\end{quote}
The page footer of Fair Play stationary quoted FDR: ``Americanism is
not, and never was, a matter of race or ancestry"~\cite{fisher}.  

Fair Play advocated a policy of \emph{dispersed relocation} for
Japanese civilians after the war, a policy in agreement with the War
Relocation Authority~\cite{sproul}.  The policy had three noteworthy
components.  (1) The policy opposed segregation into Japanese ghettos.
(2) For practical reasons, ``it is convinced that there will never be
a mass return of evacuees to the West Coast."  (3) Finally, ``the
right of loyal Japanese to come back [to the West Coast], if they so
elect, cannot be denied without a denial of all that America has
hitherto meant to racial and religious minorities, of all that it has
symbolized for the hopes of humanity"~\cite{sproul}.  During the war,
while working to change policy and public attitudes, Fair Play
held that Japanese-American rights must ultimately prevail. However,
recognizing the strong public opposition, the committee did not insist
on their immediate return.  Millikan publicly supported Fair Play's
policy of dispersed relocation.

In fact, dispersed relocation is how history played out.  According to
the \emph{Washington Post}, ``When at last the Army rescinded its exclusion
order about 57,500 evacuees moved back to their former homes in the
West Coast states. But about 51,800 settled eastward in new homes''
(March 28, 1946)~\cite{dispersed}.  Even before rescission, under
Myer's lenient policies, a significant fraction of Japanese had left
the internment camps and resettled outside California~\cite{WRA}~\cite{whelan}.



\subsection{Investigation into Pasadena's Fair Play}

The Pasadena chapter of Fair Play came under political attack in
December 1943.  Leading the attack was California State Assemblyman
Chester Gannon, who ran the California State Assembly Committee on the
Japanese Problem, which opposed the return of Japanese-Americans to
California.  Anti-Japanese sentiments were flaring up, because of
reports of Japanese riots at the Tule segregation camp the previous month.%
\footnote{
Accounts of the Tule camp events of November 1943 vary widely.  Myer
believed that ``no violence was planned''~\cite{JapLovers}.  Barbara
Takei, the author of a book on the Tule camp, has written that the
Japanese ``riots'' were actually a ``peaceful show of support'' and
dismissed the newspaper reports of rioting as ``sensationalized
tales''~\cite{Takei}.  The events of November led to the imposition of
martial law at the Tule camp, which lasted into January 1944.}
Accusing the Pasadena chapter of waging a pro-Japanese campaign,
Gannon organized hearings at the State Building in Los Angeles and
subpoenaed the entire executive committee of the Pasadena chapter.
The committee grandstanded to intimidate Fair Play and to
portray the Pasadena chapter in a bad light.  
In reaction to Gannon
committee excesses, the \emph{Los Angeles Times} published an
editorial against the committee's tendency to ``browbeat and
abuse witnesses.''  ``When they turn themselves into witch-burning
agencies\ \ldots\ they go far afield''~\cite{witch}.  A December 1943
article ``Inquisition in Los Angeles'' in \emph{Time Magazine}
described the hearings as a ``legislative romp into U.S.-Jap
baiting''~\cite{Time}.

When it was realized that Millikan had not received a subpoena, an
attorney for Gannon's committee called Millikan on the telephone for a
statement, which was read into the record of the hearings.  For the
record, ``Dr. Millikan stated that he was familiar with the statement
to this committee by Mrs. Maynard F. Thayer [Pasadena Fair Play
  chapter chair] and was in hearty accord with it''~\cite{pasadena}.

\subsection{Kuroki and Sproul speeches}

In an oral interview about the Fair Play Committee, Ruth Kingman
recalled that there were two events in 1944 that ``marked the first
real change in the attitude of the state'' of California:
the Kuroki speech at the San Francisco Commonwealth Club%
\footnote{
Today, ``the Commonwealth Club of
California is the nation's oldest and largest public affairs
forum\ \ldots\ Martin Luther King, Ronald Reagan, Bill Clinton and Bill
Gates have all given landmark speeches at the Club''~\cite{commonwealth-club}. In decades past,
it was viewed as an influential group of business and
professional men in San Francisco and northern California.} 
and the Sproul speech in southern California~\cite{kingman}.


Ben Kuroki was an American citizen of Japanese descent in the United
States Army Air Force who flew a total of fifty-eight combat missions over
Europe, North Africa, and Japan during World War II.  Monroe Deutsch,
who was on the Fair Play executive committee and the president of the
San Francisco Commonwealth Club, arranged for Kuroki to speak.  He spoke
frankly about his combat experience and the discrimination he faced.
\begin{quote}
[L]oyal Americans of Japanese descent are entitled to the
democratic rights which Jefferson propounded, Washington fought for
and Lincoln died for.

In my own case, I have almost won the battle against intolerance; I
have many close friends in the Army now -- my best friends, as I am
theirs -- where two years ago I had none. But I have by no means
completely won that battle. Especially now, after the widespread
publicity given the recent atrocity stories,%
\footnote{The ``atrocity stories'' are events at the Tule Lake camp 
before and during
a period of martial law  that lasted until
January 15, 1944.} I find prejudice once again directed against me,
and neither my uniform nor the medals which are visible proof of what
I have been through, have been able to stop it. -- Ben Kuroki, Feb 4,
1944~\cite{kuroki}
\end{quote}
Kuroki received a standing ovation. Many in the
audience were weeping.   The publicity was resoundingly positive, even
from the Hearst and McClatchy newspapers~\cite{kingman}. 


The second event was the University of California President Sproul's
speech in Los Angeles.  Katherine Kaplan, the Fair Play
executive for southern California, decided to organize a Los Angeles
chapter.  Her husband Joseph Kaplan, a physicist at UCLA and friends
with Sproul, persuaded Sproul to speak at a
luncheon in Los Angeles to launch the local chapter.  At Sproul's
suggestion, Katherine Kaplan asked Millikan to be the ``chief sponsor
of the event and to act as Master of Ceremonies.'' Millikan agreed~\cite{luncheon} \cite{kingman}.
Sproul called for increased tolerance:
\begin{quote}
The barometer of tolerance toward the evacuees is still too low
on this Coast, and the opposition is still vehement and unscrupulous.
We need your help\ \ldots\ to create an acceptance by the California
public of the enlightened way of dealing with law-abiding persons
even though they are members of an unpopular minority. --Sproul, June 29, 1944~\cite{sproul}
\end{quote}
Sproul's speech ``was considered probably the best single statement
made all during the war on the status of Japanese Americans''~\cite{kingman}.
Katherine Kaplan called it ``magnificent!'' The speech,
which was made into a pamphlet and became an authoritative
statement of Fair Play policy,
 received ``the
same degree of favorable publicity as Sergeant Kuroki got up North''~\cite{kingman}.
Thousands of copies of the pamphlet were distributed~\cite{wollenberg2012dear}.


Ruth Kingman spoke about the changes in public opinion 
after the Kuroki and Sproul speeches~\cite{kingman}.
\begin{quote}
From then on, we got opposition, but very little hate opposition. We
got a great deal of support for constitutional rights; the rights of
men in uniform; or the rights of people who were doing a job for the
country which the public would never have accepted before as being the
prerogative of anybody of Japanese ancestry. Opponents didn't talk
very often anymore. Some did, but by and large there was no further
extensive or rabid talk about ``They'll never come back!" It was, sort
of, ``Well, when they come back." -- Ruth Kingman
\end{quote}

\subsection{Millikan's participation}

Millikan's participation in the Fair Play Committee should not be
exaggerated.  There was a limit to what he could contribute.  Caltech,
which was going through major transformations during the war, needed
Millikan's decisive leadership.  Caltech became ``practically a
factory for the production of war weapons,'' building more than a
million rockets~\cite[43:564]{Millikan}~\cite[p.
  248]{millikan2020autobiography}.  At the same time, many were on
leave for war work. It was a challenging time to lead Caltech.


Despite being a busy man, Millikan contributed what he
could to Fair Play.  He was the master of ceremonies at Sproul's
speech, an event that helped turn the tide of public opinion; his
name inspired the Pasadena chapter, the most active chapter outside
the Bay Area; he helped organize the Los Angeles chapter and became a
member of the chapter's executive committee, while still serving on
the committee's central advisory board~\cite[p.~51]{luncheon}.

On September 29, 1944, not long before Roosevelt issued the
proclamation ending internment, the Pasadena chapter of Fair Play
sponsored a talk by WRA Director Myer at the local public library
auditorium.  The political factions on internment policy were
complicated and take some time to unravel.  As mentioned earlier, Myer
had become opposed to internment and pushed to dismantle the
program. However, politicians and public opinion hindered his efforts.
Fair Play supported Myer's belief in the Bill of Rights for all
citizens.  Antagonistic Hearst newspapers accused the WRA of coddling
the Japanese.  Many wanted Director Myer dismissed and replaced with a
hard-liner.

Millikan, introducing Myer to an overflowing audience, quoted parts of
Sproul's speech, emphasized the need to preserve the Bill of Rights,
and denied Hearst newspaper accusations.  Myer spoke of softening
public attitudes.  As reported by the \emph{Los Angeles Times}, ``A
changing attitude on the part of the public will make the return of
all Japanese to all sections of the country an easier job from here
on''~\cite{WRA}.  Myer won the audience over.



On December 18, 1944, the day after the proclamation from FDR that
ended internment, the chairman of the community council at the Wyoming
Heart Mountain relocation center (one of the ten Japanese internment
camps) sent Millikan a thank-you note.  The Japanese community
council, which met twice a week, formed the leadership of the Heart
Mountain center.

\begin{quote}

Dear Dr. Millikan:

\smallskip

  May we take this opportunity to thank you for your untiring efforts
in bringing the principles of this nation into proper perspective to
the people in regards to the evacuation of Japanese from the West
Coast.  We are in receipt of the good news today of the lifting of the
restriction on Japanese by the Western Defense Command.  We realize
that you played no small part in realizing this very important move.

\smallskip

--Minejiro Hayashida, Dec 18, 1944~\cite[43:931]{Millikan}
\end{quote}



The letter shows that Millikan's influence was felt directly in the
internment camp.  
Many Japanese-Americans thanked Fair Play in letters.  In 1946, after
its purpose was completed, Fair
Play was dissolved.

\subsection{Postwar scientific missions to Japan}

Millikan's wartime involvement with Japanese policy came primarily
through the Fair Play Committee.  In addition to his Fair Play
activism, Millikan also had a small indirect scientific connection
with postwar Japan.  At the end of the war, Millikan communicated with
the physicist Alan Waterman about scientific missions to Japan,
organized through the Office of Scientific Research and Development
(OSRD).  This section describes the origin of these scientific
missions and Millikan's assistance in recruiting Japanese-American
scientists and engineers.

During the war, Vannevar Bush, who reported directly to FDR, was the
head of the OSRD, a wartime agency created to conduct scientific
research for the military.  This U.S. federal government agency
coordinated almost all wartime military research and development.
Even the Manhattan Project to develop atomic bombs was initially under
the OSRD.

Within the OSRD, Bush created the Office of Field Service (OFS), and
appointed Karl Compton chief and Alan Waterman deputy.  The OFS
provided ``civilian technical expertise needed by military commands in
the field, particularly those in the Pacific.''  Its employees were
largely scientists and engineers -- especially physicists, electrical
engineers, and communications experts~\cite[p.
17]{schrader2006history}.

The OSRD and its field office were directed by renowned names in
science. Bush, who had been a dean and vice president at MIT
and an early researcher in analog computers, became a visionary of
information technology.  The physicist Karl Compton, the brother of
the Nobel laureate Arthur Compton, was president of MIT for eighteen years.
After the war, Waterman, a physicist, became the first director of
the National Science Foundation.  He is remembered today through the
prestigious Alan T. Waterman Award for scientists.

The OSRD was reorganized when the war in Europe ended.  Compton was
transferred to Manila to establish a Pacific branch of the OSRD, and
Waterman replaced Compton as OFS chief.  There were plans to expand
the Manila office to more than two hundred scientists and engineers.
However, the day after Compton arrived in Manila, on August 6, 1945,
the atomic bomb was dropped on Hiroshima.  Japan surrendered days
later, on August 14 (Victory over Japan Day or V-J Day).  The plans to
expand the Manila office were abruptly cancelled.

Instead, efforts were immediately redirected toward postwar scientific
expeditions to Japan.  According to Compton, ``Every branch of Army,
Navy, Air Force began immediately after V-J Day to get technical
investigating teams'' to evaluate Japan's scientific capabilities.
Some of the teams were to number as many as 750.  Compton and Edward
Moreland (who was Bush's successor as dean of engineering at MIT)
themselves led one of the earliest expeditions, which resulted in a
massive $850$-page scientific report~\cite[p.
17]{schrader2006history}~\cite{home1993postwar}.


It was at this moment of intense activity of the OSRD in the Pacific
when Waterman sent Millikan an urgent request. 
Waterman asked Millikan for information about ``former Japanese
students enrolled in scientific and engineering studies'' at
Caltech~\cite[31:1003]{Millikan}.  The Office of Scientific Research
had an apparent urgent need for Japanese-speaking scientists and
engineers.  Caltech's registrar and alumni office promptly produced 44
names, which Millikan forwarded to Waterman.  The Waterman-Millikan
correspondence will matter in what follows, because of the
way it was later misinterpreted.


\section{Dishonor}

\subsection{Committee on Naming and Recognition (CNR)}


A movement to have Millikan and others stripped of honors at Caltech
(such as the library named after Millikan) started during the summer
of the 2020 Black Lives Matter protests amid the widespread toppling
of statues.  The Black Scientists and Engineers at Caltech (BSEC) sent
a petition to the Caltech community on June 25,
2020~\cite[p.~36]{CNR}:
\begin{quote}
  By now we are all well-aware of the global protests calling for
  police reform following the graphic murders of Ahmaud Arbery,
  Breonna Taylor, George Floyd, and countless others at the
  hands of the officers whose supposed duty is to protect and serve. --BSEC
\end{quote}
The BSEC called on Caltech to ``reform the long-standing causes of
  racial bias which have disproportionately hurt racially minoritized
  members of the Caltech community'' and called on the Board of
  Trustees to ``rename the buildings which currently honor Nazis,
  racists, and eugenicists: Millikan, Watson, Ruddock, Chandler.''

A separate petition with many signatures was submitted by Michael Chwe
(a Caltech alum who now is a political scientist at UCLA) in July
2020.  Here are the opening sentences of Chwe's
petition~\cite[p.~39]{CNR}:
\begin{quote}
As members and friends of the Caltech community, we believe that
Caltech cannot honor individuals who actively supported and encouraged
crimes against humanity.  Therefore we call for Caltech to rename
all buildings, spaces, and programs named after Robert
A. Millikan,\,\ldots -- Chwe petition
\end{quote}

Caltech President Thomas Rosenbaum
 formed a Committee on Naming and Recognition (CNR) to examine the
 issues in the petitions.  The committee issued a
 final report on December 17, 2020, recommending ``that Caltech remove the
 names Millikan, Chandler, Gosney, Munro, Robinson, and Ruddock from
 all Institute assets and honors.''  The Caltech Board of Trustees
 endorsed the recommendation.  President Rosenbaum wrote that ``renaming
 buildings is a symbolic act, but one that has real consequences in
 creating a diverse and inclusive environment.''


A follow-up article was published in \emph{Nature} on November 10,
2021~\cite{subbaraman2021caltech}.  One student is quoted, ``I find it
important to rename the buildings just because I don't want to have
that constant reminder that the people who built this institution
didn't want me to exist.'' Another student said of the Millikan
Library, ``We shouldn't be idolizing people with horrible views of the
world.'' It is striking to find such emotionalism coming from Caltech
students and published in \emph{Nature}.

It is beyond the scope of the research here to examine all of the
claims of the CNR report and the \emph{Nature} article.  The scope 
of this section is
restricted to statements about Millikan's wartime attitudes toward
Japanese Americans.


\subsection{Bloodlines}

Anthony Platt is a retired professor of American history, public
policy, and social sciences.  He is the author of the
book \emph{Bloodlines}, published in 2006, which is the source of
false and inflammatory accusations about
Millikan~\cite{platt2015bloodlines}.%
\footnote{Cecilia E. O'Leary is listed as a coauthor of \emph{Bloodlines},
but the acknowledgments state that ``for health reasons, Cecilia was
not able to participate in the writing of this book.''  The narrative
voice is first-person singular, recounting events of Platt's life.}
This section analyzes some of the claims in \emph{Bloodlines},
especially those related to Japanese internment.  Although
false, some of these accusations became part of the official report of
Caltech's \emph{Committee on Naming and Recognition} and were among
the reasons given to strip Millikan of honors~\cite{CNR}.

The book \emph{Bloodlines} is structured around the history of
the original typescript of the Nuremberg Laws, signed by Hitler
 and enacted by Nazi Germany in
1935~\cite{platt2015bloodlines}.  The laws established the
black-white-red swastika as the national flag, declared that only
those of German or related blood were eligible for citizenship, and
prohibited marriage between German and Jew.

The original Nuremberg document fell into the hands of General George
S. Patton Jr. in Germany at the conclusion of World War II, as a spoil
of war.  When he returned to America, Patton was welcomed by
a homecoming parade along the streets of Los
Angeles on June 9, 1945.  Two days later in Pasadena, Patton
presented to Millikan the Nuremberg Laws, which were placed in the
vault of the Huntington Library for safekeeping.  At the time,
Millikan was the chairman of the board of trustees at the Huntington
Library.  A historic photograph shows Patton and Millikan standing
together under a portrait of George Washington, the document in hand.
From there, Patton traveled to Washington, D.C. to meet with
President Truman~\cite[p.~102]{platt2015bloodlines}.

A central accusation of Platt's book is that several of those who handled
the original Nuremberg Laws -- including Patton, Millikan, and some
trustees at Huntington Library -- formed an enclave of Nazi sympathizers.
Platt provocatively claimed that Millikan's ``ideological assumptions were the
same as those that guided the Nuremberg Laws.\,\ldots''%
\footnote{Platt does
concede, however, that ``Millikan resigned from the American Committee
for Democracy and Intellectual Freedom because it was not sufficiently
vigilant against Nazi and Communist agents\ldots''~\cite[p.
124]{platt2015bloodlines}. }  Platt imagined Patton and Millikan ``in
1934 shaking their heads in agreement as they read an item in the
U.S.-published \emph{Eugenical News}, reprinted from the Nazi press,
about how `large German cities' were being `literally swamped
by\,\ldots\ Jewish physicians.'\,''

Platt's accusation that Millikan was ideologically aligned with Nazi
Germany is false and takes little effort to refute.  His tale linking
Patton and Millikan to the Nazi press is purely fictional and is
inconsistent with Millikan's character.  
One convincing way to see that \emph{Bloodlines} misrepresents
Millikan is to read Millikan's own words.  Millikan's autobiography
contains several statements about
Hitler~\cite{millikan2020autobiography}.  To allow Millikan's views on
Nazism to appear without editorial selection, every single statement
is enumerated.  On page 90, Millikan asks, ``Is not the greatest
menace\ \ldots\ defined as what Mussolini, Hitler and Lenin have done
to Italy, Germany and Russia?''  On page 116, Millikan speaks of the
``degradation of Germany under Hitler.''  On page 254, Hitler is
called a gangster.  On page 256, Millikan writes, ``Had we been in the
League of Nations in 1936, prepared to do our part,\ldots he [Hitler]
could have been permanently checked.'' On page 258, Millikan wrote
that ``for the sake of overthrowing Hitler we became Russia's
ally,\,\ldots'' Finally, on page 277, Hitler is called a maniac.  Not
a single one of these statements supports the claim that Millikan was
a Nazi sympathizer.

Starting in the mid-1930s (and to a lesser extent in the 1920s),
Millikan warned America of the twin dangers of pacifism and
isolationism.  ``Long before Pearl Harbor he [Millikan] saw what was
coming and led the movement against isolationism in California''
(\emph{Los Angeles Times}, 1950)~\cite{la-barton}.  Although the \emph{Los Angeles Times}
report might be hyperbolic, as the decade progressed, he became
increasingly publicly vocal through many speeches, publications, and
radio addresses on the evils of fascism and totalitarianism.  Millikan
did not hold back.  As an admired Nobel laureate, he had enormous
influence, which he directed towards this pressing cause.  The
documentary evidence is abundant.

Millikan's evangelism against fascism had international reach.  In his
address ``India and the War'' on June 12, 1940, Millikan made an
appeal to India, following his eight-month cosmic-ray expedition to
India.  He spoke with the eloquence of a senior
statesman~\cite[67:730]{Millikan}:

\begin{quote}
In no war in history have the fundamental issues of the
struggle been made more clear either with respect to
India, the United States, or any peace loving people, for
they have been stated unmistakably in {\it Mein Kampf}.  Or, if
words are not considered sufficient evidence, and one thinks
there is any chance that the expressed purposes will not
be put into practice, the continuous stream of acts
of perfidy, barbarism, and dishonor which have accompanied
the inhuman treatment of the Jews and the successive rape
of the liberties of the adjoining little countries of
Austria, Czechoslovakia, Poland, Denmark, Norway, Holland,
Luxembourg, and Belgium make crystal clear what modern
civilization the world over can expect from the triumph
of the Nazis.

Between the ideology of conquest and that of rational,
peaceful change there is no possible compromise.\,\ldots\
This is the time for every American and every Indian
and every peace loving man everywhere to exert every
ounce of influence he can to prevent the destruction
of civilization and the return of the horrible tyrannies
and despotisms that have cursed mankind through all history.
-- Millikan, June 12, 1940
\end{quote}


\subsection{Bloodlines on internment policy}\label{sec:blood-internment}

This subsection analyzes statements from the book \emph{Bloodlines}
about Millikan's activism for Japanese-American rights.
Millikan's activism in the Fair Play Committee for the
protection of Japanese rights does not square with \emph{Bloodlines's} claim
that Millikan was ideologically a Nazi.  According to \emph{Bloodlines},
``On the `Japanese Problem' during World War II, Millikan took a
more complicated, but ultimately opportunistic
position''~\cite[p.~126]{platt2015bloodlines}.  Millikan's position was
not complicated: his support of the Fair Play Committee was unwavering
throughout the war.  Millikan was not opportunistic in the sense of
seeking self-gain; nor did his actions lack ethical principle.  Most
criticisms by Platt of Millikan on the Japanese issue are non-specific
and suggest that Platt held the entire Fair Play Committee in low
regard.

Two sentences from \emph{Bloodlines} are particularly relevant.
Platt wrote~\cite[p.~126]{platt2015bloodlines}:
\begin{quote} In 1943, Millikan told counsel for an Assembly committee
investigating the dangers of treason that he favored dispersal throughout
the country of California's Japanese at the end of the war. -- Platt
\end{quote}
The sentence refers to Millikan's telephone conversation during
Assemblyman Gannon's committee hearings, as discussed earlier.  The
sentence is a peculiar way to describe the Gannon hearings because
of all that it leaves out --
that the Gannon committee
and Fair Play were political adversaries, that Fair Play was the
target of the investigation,
that
Millikan was called up because of his support of
Fair Play,
that the committee had the tendency to browbeat witnesses, 
and that \emph{Time Magazine} called the hearings
an inquisition.

The mention of ``dispersal'' in Platt's sentence is a reference to
Fair Play's policy of dispersed relocation, which is discussed in
detail above.  The policy declared a Japanese right of return to
California as soon as the exclusion order was rescinded, while
recognizing practical reasons for voluntary partial dispersal outside
California.  Millikan endorsed this policy.

The sentence from \emph{Bloodlines} makes Fair Play's position sound
sinister by failing to mention its insistence on Japanese-American
rights and by setting the encounter at a ``committee investigating the
dangers of treason.''  The insertion of the dangers of treason into
the context is a red herring: Fair Play's policy was entirely
unrelated to dangers of treason.  Indeed, Fair Play waged a public
relations campaign against false but widespread accusations of
Japanese-American disloyalty.  From its very first press release, Fair
Play stated that it would be un-American to transfer a ``distrust of
the Japanese Government'' ``to all persons of Japanese race.'' The CNR
copied verbatim Platt's sentence on the Gannon investigation into its
report, without citing \emph{Bloodlines} as its source.


\smallskip

Here is the second relevant sentence from \emph{Bloodlines}~\cite[p.
126]{platt2015bloodlines}:
\begin{quote}
After the war ended, Millikan did not hesitate to turn over to
military intelligence the names and known address of all students of
Japanese background who had studied at Caltech between 1929 and 1944.
-- Platt
\end{quote}
The sentence has an ominous feel to it.  Millikan almost sounds like a
wartime collaborator with military intelligence against Japanese Americans, except
that cannot be, because of the timing after the war.  Platt did
not give context, but this article has supplied extensive context in the
section on postwar scientific missions to Japan. Platt
left out essential details: the request came from a
long-time acquaintance, the physicist Waterman; the request asked specifically
for scientists and engineers; the request came from the Office of
Scientific Research and Development, which employed many scientists and
engineers; the request was written on V-J day and was urgent.

The CNR copied Platt's sentence into its report, without citing
\emph{Bloodlines} as its source. The sentence was copied verbatim, except for a tiny
but significant change.  Instead of writing, ``after the war ended,''
the CNR wrote ``ca. 1945.''  Crucially, in the CNR report, it becomes
possible to interpret the Millikan's action as occurring before the end
of the war.

A game of telephones is in play here, where a message becomes
increasingly garbled with each repetition.  The original context of
postwar scientific missions to Japan is described in detail earlier.
Platt leaves out essential context and makes the Waterman-Millikan
correspondence sound ominous but strangely anachronistic, because of
the timing after the war.  The CNR report modified the date so that
the exchange was no longer unambiguously after the war.

The final stage of the game of telephones is provided by Nidhi Subbaraman,
writing for \emph{Nature}.  The year has been altered from 1945 to 1942, from
the end of the war to the beginning of internment. No trace remains of
the original historical context of postwar scientific missions to
Japan.  The Waterman-Millikan correspondence on scientific recruitment
was falsely reported the following way in
\emph{Nature}~\cite{subbaraman2021caltech}.

\begin{quote}
During the Second World War, as the United States began a
nationwide effort to imprison civilians living in the country
who had Japanese ancestry, Millikan collected the names and
addresses of Japanese students who had studied at Caltech
in the previous two decades and passed the list to the US military.
-- Subbaraman, Nature 2021
\end{quote}
This is simply false.  From the syntax of Sabbaraman's sentence, it is
recognized as being derived from \emph{Bloodlines} as modified by the
CNR report.  However, the game of telephones has fully corrupted the
meaning.  The corruption progressed from
original sources, to Platt, to the CNR report, and finally to Nature.
At each stage, the meaning changed unjustifiably  in the same direction:
always to injure Millikan's
reputation and never to bring him favor.

\section{Eugenics in 1938}

According to Caltech President Thomas Rosenbaum,
``The most intense concerns at Caltech center on Robert A. Millikan,''
because 
he ``lent his name and his prestige to a morally
reprehensible eugenics movement that already had been
discredited scientifically during his time''~\cite{rosenbaum}.
This section analyzes the evidence in the CNR report in
support of the claim that the eugenics movement had been
scientifically discredited by 1938~\cite{CNR}.%
\footnote{
This section continues to refer to the 
Committee on Naming and Recognition (CNR) as the CNR or
simply \emph{the committee}.  The \emph{report} refers to the CNR
report.  HBF refers to the Human Betterment Foundation.
}  The discussion is restricted in
scope to scientific claims and does not treat the moral and political
dimensions of the eugenics movement.  The word \emph{eugenics} evokes
many connotations; a starting point is 
the Oxford English
Dictionary definition of eugenics:
``the study of how to arrange
reproduction within a human population to increase the occurrence of
heritable characteristics regarded as desirable.''
From this starting point, the definition diverges in many
directions~\cite[p.~44]{wellerstein}~\cite{koch}.  


The CNR report presents statements from three of Millikan's scientific
contemporaries (Lancelot Hogben, Hermann Joseph Muller, and Thomas
Hunt Morgan) to establish by their authority that eugenics had
``fallen into disrepute'' by the late 1930s~\cite{CNR}. 

\subsection{Hogben}

The first of the committee's three authorities against eugenics was the
medical statistician Lancelot Hogben (and author of the 1936
best-seller \emph{Mathematics for the Million}).  According to the CNR
report, ``In 1931, geneticist Lancelot Hogben declared that `all the
verifiable data eugenicists had accumulated on the inheritance of
mental traits could ``be written on the back of a postage
stamp''\relax'\relax''  \cite{spiro}~\cite{CNR}. 
In truth, Hogben declared no such thing.  In fact, what Hogben actually wrote
was that
\begin{quote}
all existing and genuine knowledge about the way in which the
physical characteristics of human communities are related to their
cultural capabilities can be written on the back of a postage
stamp.\,\ldots\ there is as yet no biological knowledge bearing on the
social capabilities of different `races'\ \ldots\ --Hogben (from Hogben's
1937 preface to \emph{Half-caste} \cite{half-caste} and reprinted in \cite[p.~47]{hogben})
\end{quote}
In the preface to the book \emph{Half-caste}, Hogben built an
argument that children of mixed marriages should be afforded the same
cultural advantages as other children.  The subject matter was not
eugenics, and
it was reckless scholarship to alter Hogben's statement to make 
it appear to be.

Hogben himself wrote more than a postage stamp's worth about eugenics,
which is the subject
of the 
entire last chapter 
of his book \emph{Genetic Principles in Medicine and Social Sciences}
(1931)~\cite{hogben-genetic}.  Hogben's principled stance of scientific
detachment kept him from making policy recommendations.  
Nonetheless, despite the detachment, Hogben most certainly
did not view eugenics as scientifically unfeasible.

He noted, ``Eugenics was
defined by [Francis] Galton as the study of agencies under social control which
may improve or impair the racial qualities of future generations.
With such a proposal it is difficult to see that any reasonable person
would disagree''~\cite[pp.~209--210]{hogben-genetic}.

Hogben wrote that he ``would prefer to use the term `genetic therapy'
for the legitimate province of applied human genetics,'' because of
negative political associations that the term \emph{eugenics} had
acquired by 1931.  Nevertheless, eugenists
have begun ``to write with greater caution'' in the past two decades.
``[W]e can agree about certain disorders which practically all
comparatively healthy people would wish to remove''~\cite[p.~213]{hogben-genetic}.

In California, eugenic practice took the form of a large sterilization
program.  The state government ran several hospitals for the care of
those with mental illness or intellectual disabilities.  During the
years 1909--1979, over twenty thousand patients at those hospitals
were sterilized~\cite{stern}.  The ``operations were ordered at the
discretion of the hospital superintendents,'' as authorized by
California law~\cite{wellerstein}.  About three-quarters of
sterilizations in California in 1936 were performed by request or
consent of the patient or
guardian~\cite[Popenoe:18:01]{hbf}~\cite{wellerstein}.%
\footnote{ Calculations of consent rates were based on signed consent
  forms.  However, standards of written consent in the 1930s compare
  infavorably with ethical and legal standards of informed medical
  consent today.  In some cases, hospital release was contingent on
  consent to sterilization~\cite{stern-nation}.  In an analysis of the
  reasons for lack of consent among Spanish-surnamed patients during
  the years 1935--1944, the most frequent reason was that ``no
  consenter was available''~\cite[p.~148]{lira}.}
\footnote{A strong distinction should be made between voluntary and
  involuntary sterilization.  Worldwide today, almost half of all
  ``women of reproductive age (or their partners) are contraceptive
  users''~\cite{UN}.  According to data from a recent study in Lancet,
  sterilization is still the most common form of birth control in the
  world with prevalence around 26\%~\cite{GBD}.  Decades of
  sterilization research (including Paul Popenoe's at the HBF) have
  shaped our understanding of worldwide contraceptive practice today.}

About sterilization laws in California, Hogben wrote, ``As a
precautionary measure, there do not seem to be any strong arguments
against Californian laws on administrative or clinical grounds.  So
long as the advocates of such a policy do not claim that it is more
than a precautionary measure, it would be unreasonable to criticise
the results achieved so far,\,\ldots''~\cite[pp.~207--208]{hogben-genetic}.

In summary, the committee used a fabricated ``postage stamp'' quotation to
attribute views to Hogben that he did not hold.  In truth, Hogben
claimed it is ``difficult to see that any reasonable person would
disagree'' with eugenics in Galton's sense of the word.

\subsection{Muller}
We turn to Caltech's second scientific witness against eugenics in the 1930s.
The CNR report states that
\begin{quote}
  Also in 1932, future Nobel laureate H.J. Muller denounced eugenics as an
  ``unrealistic, ineffective, and anachronistic pseudoscience'' in his paper
  ``The Dominance of Economics Over Eugenics.'' \cite[p.~13]{CNR}\ \cite{spiro}.
\end{quote}
This quotation is also a
fabrication.  Muller made no such denunciation. On the contrary,
that very paper declares, ``That genetic imbeciles should
be sterilized is of course unquestionable\ldots''~\cite[p.~40]{dominance}.

What did Muller mean by the dominance of economics over eugenics?  His
``dominance'' paper was an address to the Third International Congress
of Eugenics in 1932.  Muller had become a devoted Marxist. The address
was a Marxist critique of capitalism, where he proclaimed an
``impending revolution in our economic system''~\cite{dominance}.
Muller's New York audience was probably bewildered by his heavy-handed
ideology.  
``Muller was
convinced that his eugenic ideas could only flourish in a socialistic
state''~\cite[p.~351]{allen}. Marxism and eugenics went hand in hand:
Marxism to fix the environment and eugenics to fix heredity.
Socialism claimed dominance over eugenics because revolution was nigh
at hand.  ``In place of the economic conditions imposed by the class
struggle, entirely new conditions will be substituted.\,\ldots\ True
eugenics can then first come into its own,\,\ldots\ Thus it is up to
us, if we want eugenics that functions, to work for it in the only way
now practicable, by first turning our hand to help throw over the
incubus of the old, outworn society''~\cite{dominance}.

True to his beliefs, Muller, who was born in New York, established a
lab in the Soviet Union.  His timing could hardly have been worse.
``It was not long before he realized that the conditions for the
development of a Bolshevik eugenics were less promising than he had
assumed''~\cite[p.~579]{paul}.  Muller's 1936 book, \emph{Out of
  the Night,} did not please Joseph
Stalin, and he fled the Soviet Union under inauspicious circumstances.

According to the historian Garland E. Allen, Muller's book ``occupies
a significant place in the history of eugenic writing''~\cite{allen}.
Muller wrote, ``Thus we see that only the eugenics of the new society,
freed of the traditions of caste, of slavery, of colonialism, can be a
thorough-going and true
eugenics''~\cite[p.~150]{muller-out}.\footnote{Magnus Hirschfeld held
  eugenic views similar to those of Muller and quoted this particular
  sentence in his 1938 book \emph{Racism}~\cite[p.~174]{hirschfeld}.}
\begin{quote}
  In time to come, the best thought of the race will necessarily be
  focused on the problems of evolution -- not of the evolution gone
  by, but of the evolution still to come -- and on the working out of
  genetic methods, eugenic ideals, yes, on the invention of new
  characteristics, organs, and biological systems that will work out
  to further the interests, the happiness, the glory of the god-like
  beings whose meagre foreshadowings we present ailing creatures
  are. -- H. J. Muller, 1936 ~\cite[p.~156]{muller-out}
\end{quote}


The geneticist Elof Carlson, who received his PhD under Muller's
mentorship, discovered evidence that Muller had suggested in private that 
Stalinesque coercion
might be used should voluntary programs fail.  In Carlson's words,
Muller envisioned that eugenic ``controls might be imposed as a second
step, just as, after the Soviet Revolution, the land was given to the
peasants, but when agriculture remained as backward as under the
czars, Stalin had to impose a strict control over the land,
collectivizing the farms,\,\ldots''~\cite[p.~186]{Carlson-gene}.


Hogben, Muller, and Caltech geneticist Dobzhansky all signed the
``Geneticists' Manifesto'' at the 1939 International Congress of
Genetics.  An article about Ronald Fisher's involvement in
eugenics states that
\begin{quote}
this document was signed by 23 leading geneticists,
including some with strongly left-wing political views like
J.B.S. Haldane, H.J. Muller and Lancelot Hogben. It started with the
question ``How could the world's population be improved most
effectively genetically?'' They went on to say that ``the raising of
the level of the average of the population nearly to that of the
highest now existing in isolated individuals\ \ldots\ would, as far as
purely genetic considerations are concerned, be possible within a
comparatively small number of generations''. This goes far beyond the
proposal of Fisher's,\,\ldots\ and shows that eugenic ideas
were widely held across the political spectrum at the time (see Paul
1984~\cite{paul} for further discussion).-- Bodmer et al. 2021~\cite{bodmer}
\end{quote}

The manifesto called for a process of ``conscious selection'' to
replace natural selection in human evolution.  ``The most important
genetic characteristics'' for conscious selection should be health,
intelligence, and prosocial behavior~\cite{crew}.  
Some -- including Muller,
J.B.S. Haldane, and Julian Huxley -- ``continued to argue the case for
eugenics into the 1960s\,\ldots'' \cite[p.~589]{paul}.  Just months
before his death in 1967, in his last public address, Muller proposed
gene selection to enhance cooperative behavior among
humans~\cite[p.~352]{allen}~\cite{muller67}.

\subsection{Morgan}

The committee's final scientific witness against eugenics was Thomas H. Morgan.
The CNR report states, ``By this time [1937], many geneticists -- including
Nobel laureate and Caltech professor Thomas H. Morgan -- had already denounced
eugenics for its lack of scientific merit''~\cite[p.~21]{CNR}.
``In 1925, in his book \emph{Evolution and Genetics}, Morgan criticized
eugenics for its interpretation of `feeble-mindedness' and its insistence
on the genetic basis for such characterological traits''~\cite[p.~13]{CNR}.

Indeed, Morgan's book does contend that based on available evidence it
would be ``extravagant to pretend to claim that there is a single
Mendelian factor for this condition'' of feeble-mindedness
\cite[p.~201]{morgan}.  He stated that until certain questions ``are
better understood it is impossible to know how far observed
differences are innate and how far acquired'' \cite[p.~200]{morgan}.
On display here is Morgan's modesty in the face of enduring questions
about heritability.

He then expressed his opinions more fully.
\begin{quote}
Lest it appear from what has been said that I have too little faith in
the importance of breeding for mental superiority I should like to add
that I am inclined to think that there are considerable individual
differences in man that are probably strictly genetic, even though I
insist that at present there is for this no real scientific evidence
of the kind that we are familiar with in other animals and in plants.
I will even venture to go so far as to suppose that the average of the
human race might be improved by eliminating a few of the extreme
disorders, however they may have arisen.  In fact, this is attempted
at present on a somewhat extensive scale by the segregation into
asylums of the insane and feeble-minded.  I should hesitate to
recommend the incarceration of all the relatives if the character is
suspected of being recessive, or of their children if a
dominant.\,\ldots\ How long and how extensively this casual isolation of
adults would have to go on to produce any considerable decrease in
defectives, no informed person would, I should think be willing to
state. --  T. H. Morgan\ \cite[p.~206]{morgan}.
\end{quote}
In summary, in 1925 Morgan tentatively believed that there were significant individual
differences that ``are probably strictly genetic.'' He did not condemn
eugenic practices in the asylums.  However, he hesitated to
recommend extending incarceration to relatives.

Morgan agreed to join the board of the Human Betterment Foundation
(HBF), a eugenic research and advocacy group based in Pasadena,
California.%
\footnote{For unexplained reasons, the
  committee chose to judge Morgan and Millikan by
  entirely different standards: one man deemed good and the other evil.}
In a letter to board member William B. Munro on March 17, 1942, the 
founder Ezra S. Gosney wrote,
\begin{quote}
Dr. T. H. Morgan has promised me that he would serve.  I was very glad
to find that he approves our $8$-page pamphlet, and in reply to my
request for constructive criticism he said ``I would not change a word
in the pamphlet, it is all right.'' He has manifested more interest in
our work than I had expected him to show. --Gosney, 1942~\cite[01:04:26]{hbf}
\end{quote}
The eight-page pamphlet refers to \emph{Human Sterilization Today},
authored by Gosney in 1937~\cite{gosney}.  Morgan was voted
unanimously onto the board, but because of Gosney's death later that
year, Morgan's participation was cut short.  

The next section will present Millikan's views on eugenics.
Morgan surpassed Millikan in eugenic piety by approving the pamphlet.
There is no evidence that Millikan ever read the pamphlet or endorsed
it.  Board members were free to disagree: board membership did not
imply endorsement of the pamphlet.\footnote{According to Gosney,
``Dr. Munro was \emph{against} our first print [of \emph{Human
  Sterilization Today}], Dr. Robert Freeman
favored\ \ldots''~\cite[01:01:23--25]{hbf}.  Munro's name appears in
the pamphlet as a board member, \emph{despite his opposition} to the
publication.}

\subsection{Millikan}

The eugenics movement was a \emph{big tent} that
transcended national boundaries and political
ideologies~\cite{bashford}~\cite{stepan1991hour}.  In 1938, Haldane
felt that these questions ``cut right across the usual political
divisions''~\cite[p.~9]{Haldane}.  ``Many eugenicists had very low
regard for one [an]other and routinely disagreed with others in the
community''~\cite[p.~45]{wellerstein}.  These squabbles did not signal
the end of the movement.
Further evidence could be presented that the scientific community was
not united against the feasibility of eugenic interventions in 1938.
However, the committee's case has been amply refuted by its own
witnesses. 

Where did Millikan fit into the eugenics movement?  He was a bit
player.  His name does not appear in the definitive histories of
eugenics, and biographies of Millikan do not mention eugenics.
Caltech emeritus professor Daniel J. Kevles, who is a leading
authority on both Millikan and eugenics, did not mention Millikan in
his history of eugenics~\cite{kevles}.  Overwhelmingly, Millikan's
most significant contribution to the biological sciences was his part
in establishing Caltech's division of biology \cite{kay}.  Caltech was
his legacy.

The following context is relevant to understanding Millikan's
relationship with the California eugenics movement.
\begin{enumerate}
\item In 1938, Millikan became a board member of the Human Betterment
Foundation.%
\footnote{The CNR report states that Millikan joined in 1937, but the
  correct date is 1938.  He replaced the banker Henry Mauris Robinson
  on the board, who died on November 3, 1937.  Gosney wrote a letter
  on January 19, 1938 describing plans to ask Millikan to join: ``We
  could promise him [Millikan] that it would require a minimum of his
  time''~\cite[Gosney:08:13:17]{hbf}.  } He did not attend the annual
board meetings.  Millikan's non-participation is documented in the
form of signed proxy-vote slips that are in the Caltech archives for
those years that the board met~\cite{hbf}.  By the time Millikan
joined, the foundation was dying.  Gosney was old, and his health
was failing.  Research projects at HBF had
come to an end by May 1937, when Paul Popenoe resigned from the
organization~\cite[Reid:08:14:49]{hbf}.  In its final stage, the
foundation's only significant activities were the management of its
real estate assets and the distribution of old pamphlets.%
\footnote{The foundation owned a six-story loft building in Los
  Angeles, business property in San Bernardino, and ranch
  property~\cite[Castle:08:14:60]{hbf}.}
\item 
The foundation closed down after Gosney's death in 1942,
and its assets were eventually donated to Caltech.
  In consultation with Gosney's daughter, it was Millikan who
  redirected the funds \emph{away} from sterilization research~\cite{beadle}.
\item 
Millikan and other public intellectuals of the
  1930s sponsored an enormous number of civic organizations, events,
  and causes.  The vast array of causes that Millikan endorsed has never
  been catalogued.
   His \emph{Nachlass}, which is more than 125,000 pages,
  records many of them~\cite{Millikan}.
  For Millikan, the HBF was a drop in the ocean.
Millikan's permission to use his
  name was often no more than words to the effect 
 ``I am willing to agree to the use of my name as you suggest''~\cite[45:433]{Millikan}.
Sometimes these endorsements were recognized by a seat on the board of
these organizations.  These honorary positions do not generally mean that
Millikan governed the organizations.
Millikan viewed this system of
  patronage as wholesome, but it was imperfect at best.
  He once complained that ``executive committees over which the
  sponsors have no control make commitments and take actions which the
  sponsors might warmly oppose''~\cite[41:183]{Millikan}.  Privately, Millikan murmured that
  John Dewey and Albert Einstein endorsed causes they knew little about, yet
  Millikan was also guilty.%
\footnote{Voluminous FBI files catalogue many of Einstein's
  endorsements.  The comparison between Einstein and Millikan is
  instructive.  Should Einstein's endorsements and offensive travel
  diaries be evaluated today by the same standard as Millikan's
  endorsements and mean-spirited private remarks~\cite{einstein}? Many
  (including the author) might find this standard too severe.}
Millikan grumbled, ``One of my friends tells me that my name is being
so used\ \ldots\ as to let the public think that I am thoroughly
familiar with its activities and am one of its sponsors, when the
facts are that I have known nothing whatever about the management or
the program\ \ldots\ for something like twenty years.\,\ldots'' ``I
cannot look up even a tenth part of the institutions seeking my
sponsorship''~\cite[42:359]{Millikan}.
\item Because the  committee did not have access to Millikan's own
  statements on eugenics, its report was largely speculative.
  The committee relied on private words of praise (phrases such as
  ``magnificent job'') from Millikan to Gosney to infer that
  Millikan's beliefs were the same as Gosney's~\cite[p.~16]{CNR}.%
\footnote{
In a 1945
letter to Clarence Gamble, the heir of the Proctor and Gamble fortune, 
Millikan referred to the establishment of the Human Betterment Foundation as
``epoch-making''~\cite[28:441]{Millikan}.  } However, it is rash to
presume that Millikan believed what
Gosney did.  
These were the words of appreciation from a university
executive to a philanthropist, written
in response to Gosney's ``generous gift'' and ``philanthropic
enterprises.''  Gosney was wealthy and ``supported countless
charities and educational programs''
in addition to the HBF~\cite[1.3:119]{troendle}.
\item After the CNR report was issued, two statements about eugenics
  in Millikan's own words surfaced.  These direct statements from
  Millikan supersede the committee's speculations.  His first statement
  was made to a crowded auditorium during an event organized by the
  Ebell of Los Angeles women's organization in 1925. Millikan spoke
  out in strong opposition to eugenics, denouncing the race
  degeneration theories of Albert Edward Wiggam and Lothrop
  Stoddard~\cite{wiggam}~\cite{stoddard1922revolt}.  Millikan
  maintained that ``conscious selection by eugenics\ldots\ has not been
  done nor do I think it possible to do on any large scale throughout
  the world.\,\ldots\ We can't control the germ plasm but we can
  control education'' and consequently, education was the ``supreme
  problem'' and a ``great duty''~\cite{millikan-wiggam}.
\item Millikan's second statement on eugenics appeared fourteen years
  later in a 1939 article \emph{Science and the World of Tomorrow},
  which contained his forecast of how science might change ``life in
  America fifty or a hundred years hence''~\cite{millikan-tomorrow}.
  He readily admitted that ``the possibility that something so
  completely foreign to my thinking may happen as to make any
  prognosis that I may hazard now look ridiculous in the years to
  come,\,\ldots'' This article was written not long after he joined
  the Human Betterment Foundation (HBF).  If we are looking for the
  message that Millikan might have delivered to the HBF had he ever
  attended a board meeting, it is here that we should look.  He wrote,
\begin{quote} I have no doubt that in the field of public
health the control of disease, the cessation of the continuous
reproduction of the unfit, etc., big advances will be made, but
here I am not a competent witness, and I find on the whole those
who are the most competent and informed the most conservative.%
-- Millikan, 1939~\cite{millikan-tomorrow}.
\end{quote}
There is no mention of specific methods such as sterilization.  There
is also no suggestion of the use of coercion.  He does not endorse
current practices, but looks to the indefinite future;
``advances will be made'' in the coming fifty or hundred years.
He is cautious.  He humbly professes his lack of expertise.  This
brief statement is all we have; no others have surfaced.  His statement falls at the conservative
end of the range of views held by Caltech geneticists of that time.%
\footnote{
Morgan, who served alongside Millikan for many years on Caltech's
executive council, held views that are documented above.  
Dobzhansky signed the 1939 manifesto, which proclaimed that a genetic
transformation of the human population might be achieved ``within a 
comparatively small number of generations.'' By 1946, he had
revised his time estimates; a sterilization program may ``take centuries or even
  millennia.\,\ldots\ It is, perhaps, not too selfish to say that
  posterity should be allowed to tackle its own problems and to hope
  that it may have better means for doing so than we
  have''~\cite{dobzhansky}.  Decades later in 1965, Alfred Sturtevant contended that
  eugenics was fraught with uncertainties and
  difficulties~\cite[pp.~130--132]{sturtevant}~\cite{lewis}.}
As this section has documented,
Millikan held milder views than the committee's
three witnesses against him.  
\end{enumerate}

\section{Conclusions}

Millikan was one of the greatest experimental physicists in the world
during the early decades of the twentieth century.  He made
fundamental contributions to the isolation and measurement of the
electron charge, the experimental verification of Einstein's
photoelectric equation, the measurement of Avogadro's number and
Planck's constant, the study of spectral lines of ionized atoms, and
the understanding of cosmic rays.  He spurred the growth of American
science not only through his many contacts with Bell Labs and the
almost countless number of graduate students he supervised but also
through his best-selling physics textbook.  At the head of Caltech
during the first twenty-four years of its existence, Millikan oversaw
its rapid maturation into one of the outstanding technical
institutions of the world.

The CNR report's account of Millikan's scientific achievements was
deficient; this article has aimed to remedy that deficiency.  About
Millikan's famous oil drop experiment, the Caltech report seems to
have no institutional memory.  The report quoted only the first
sentence from an \emph{Encyclopedia Britannica} article about the
experiment.  Millikan and Fletcher measured the elementary charge and
electron mass.  As the number of fundamental physical constants is
limited, the measurement is a rare achievement in the history of
science.  Concerning the photoelectric effect, the Caltech report
exhibited a similar lack of institutional memory, by going no further
than the one-sentence banner statement from NobelPrize.org.  In brief,
Millikan's greatest scientific achievements were compressed to two
stock sentences.  By contrast, the CNR members care deeply about their
own reputations by including seven pages of their bios in the report.



This article has made an extended commentary on Platt's ideas because
of his ultimate influence on the decision to strip Millikan of honors.
His book \emph{Bloodlines} is an unreliable source for Millikan
scholarship.  Although Platt is more a storyteller than a careful
historian when it comes to Millikan, his fingerprints are everywhere.
Platt was cited five times in Michael Hiltzik's \emph{Los Angeles
  Times} article, ``Caltech faces reckoning,'' which made the petition
against Millikan known to a large public~\cite{hiltzik}.  The
\emph{Los Angeles Times} published Platt's false accusation that a
quota existed at Caltech during Millikan's tenure ``allowing for the
appointment of only one Jewish full-time faculty member per year.''
More than two full pages of the archivist's
eight-page report to the committee are direct quotations from Platt's
book~\cite{collopy}.  The committee itself relies heavily on
\emph{Bloodlines} (especially the error-ridden pages 124--126) but
misattributes the research to others.  Michael Chwe's presentation to
the committee quoted Platt's book four times~\cite{chwe}.  Like Platt,
Chwe imagined far-fetched associations between Millikan and Nazi
Germany.  Platt's accusations that ``California's elite'' had
``fascist sympathies'' set the tone of moral condemnation.  Through an
escalation of rhetoric, Millikan's thoughts and deeds came to be seen
as ``reprehensible'' (Thomas Rosenbaum), ``horrible'' (Daniel Mukasa,
Caltech President of BSEC), and ``crimes against humanity'' (Michael
Chwe).  The historian Kirt von Daacke has called for us collectively
to atone for this history.

Regarding the history of sterilization in California, the science
historian Alex Wellerstein has stated, ``We do not find Nazis
in California mental health institutions''~\cite{wellerstein}.
The alleged association between Millikan and
Nazi Germany is refuted by Millikan's 
speeches, publications, and radio addresses.  

Throughout World War II, Millikan advocated for the rights of Japanese
Americans.  At a time when public opinion had turned overwhelmingly
against Japanese Americans, Greta and Robert Millikan were both compassionate
toward their Japanese friends and acquaintances.  When
she received a letter from her friend Elizabeth Ozawa, who was in a
detention camp at Tulare, Greta wrote in her diary that ``things are far
from right -- we must keep alert and busy on our Fair Play
Committee'' \cite[80:132]{Millikan}.  Robert Millikan wrote to Hori
that ``your letter touched me deeply'' and to Kido that he had been
doing all he could to assist ``the many who are likely to suffer
unjustly in this difficult situation.''
For Millikan, who regularly executed his plans through a broad
network of committees, it was natural to reach out to a committee that
shared his views on Japanese-American rights.  He became a vice
president of Fair Play, which worked tirelessly to change public
opinion on Japanese-American issues and had a significant impact on
public policy.

\emph{Bloodlines} was the only source about Japanese internment in
Caltech's report.  On Japanese policy, Platt omitted key parts of
Fair Play's recommendation of dispersed relocation, which fully
recognized the right of Japanese Americans to resettle after the war
wherever in the United States they pleased.  He falsely described the
physicist Waterman's recruitment of Japanese scientists for postwar
expeditions to Japan, making Millikan seem to be a military informant
against Japanese Americans.

Remarkably, the committee selected the worst parts from the worst
secondary source, without consulting any primary sources on Japanese
internment.  Here, \emph{worst} is meant both in the sense of being
historically inaccurate and of being hostile toward Millikan.  The
report did not mention Millikan's significant participation in Fair
Play.  The report severely misrepresented the confrontation between State
Assemblyman Gannon and Millikan; in fact, it was Gannon who opposed
Japanese-American rights and Millikan who supported their rights.
Finding itself at the final stage of a game of telephones,
\emph{Nature} falsely accused Millikan of colluding in a military
roundup of Japanese for imprisonment at the beginning of the war.  Not
only is this accusation false, it is a complete reversal of historical
fact.  Millikan, who was one of a small minority who actively promoted
Japanese-American rights during the war, received the Kansha Award,
which recognizes ``individuals who aided Japanese Americans during
World War II''~\cite{kansha}.


After a discussion of Millikan's views on Japanese
internment, this article turned to the California eugenics movement.
According to Rosenbaum, Caltech's most intense concern was Millikan's
association with a eugenics movement that had already been
discredited.  The committee expanded the accusation, claiming he
``failed to perform the due diligence'' and was ``derelict in this
duty'' to ensure that the HBF had its ``science right''~\cite{CNR}.

Alas, it was the committee that failed to perform due diligence.  The
committee cited three authorities to prove that the eugenics movement
had been discredited scientifically by 1938.  In a dramatic
denouement, all three of the committee's authorities had various
degrees of eugenic involvement at that time.  Two of them signed the
pro-eugenic Geneticists' Manifesto in 1939, which proclaimed, ``The
truth that both environment and heredity constitute dominating and
inescapable complementary factors in human wellbeing, but factors both
of which are under the potential control of man and admit of unlimited
and interdependent progress''~\cite{crew}.

In the committee's view, ``The hereditary nature of human behavior and
character had fallen into disrepute within various quarters\ldots'' by
1937~\cite{CNR}.  Millikan, who believed that education could be
controlled but not the germ plasm, did not hold strong hereditarian
opinions.  The committee was quick to condemn the
hereditarians of that era,
but it would be rash to pass sentence, without grounding that judgment
in recent science.  To summarize in a few
sentences, the statistical concept of heritability provides a scale
that quantifies environmental and genetic influence.  Evidence in
support of rather high heritability of intelligence and various human
traits is presented in~\cite{plomin}~\cite{warne}~\cite{haier}.
Richard J. Haier, who pioneered the use of neuroimaging in
intelligence research, has written, ``Although the full role of genes
is not yet known, the evidence for major genetic involvement in
intelligence is overwhelming''~\cite{haier}.

Nobody proposes a return to the California sterilization practices of
the 1930s.  According to Wellerstein, sterilization rates in
California declined sharply in the early 1950s for bureaucratic
rather than moral reasons.  
``No one took credit for killing the practice, and no one at the time
appears to have noticed that it had ended.''  ``The horror we attach
to sterilizations today, and to eugenics in general, did not become
widespread until the 1970s, with the rise of interest in patient
autonomy, women's rights,'' among other
reasons~\cite[pp.~49--51]{wellerstein}.  During earlier decades, the
largest institutional force in moral opposition to eugenics had been
the Roman Catholic Church~\cite{durst}.  However, today the moral
outrage at Caltech has grown out of the Black Lives Matter movement
and was directed toward the cause of ``dismantling Caltech's legacy of
white supremacy''~\cite{sam}~\cite{CNR}.  Technologies have also
changed in irreversible ways since the 1930s, precluding a return to
the past: birth control and genetic engineering have advanced far
beyond the capabilities of the sterilization era~\cite{metzl}.

The committee depicted Millikan as a morally deplorable man and relied
on fabricated quotations to buttress its case.  To be clear, the
committee was not the original source of the fabricated quotations it
used.  Nevertheless, the words ring false.
When the fabrications are corrected,
the central accusations against Millikan crumble.
It is telling that the evidence unravels so
sensationally.  Although Rosenbaum claimed that the committee reached
its conclusions ``by close reading of primary
sources,'' his statement is not credible~\cite{rosenbaum}.

\bigskip
\centerline{\rule{3em}{0.333ex}}
\bigskip

This article has focused on Millikan, but the larger message of the
CNR report is diversity.  The word \emph{diversity} and its
inflections are used $78$ times in the $77$-page CNR report, occurring
as many as nine times per page.  The report, which is hosted by the
Caltech diversity website, makes repeated reminders that Caltech has
an ``ongoing effort to forge a diverse and inclusive community.''

The CNR was part of a package of diversity reforms at Caltech,
which included a new admissions advisory committee~\cite[p.~46]{CNR}.
 Until recently, Caltech was unique
among the most elite.  The Pulitzer Prize winning journalist Daniel
Golden has written on admission practices at elite American
universities.  Not long ago, Caltech boasted that on matters of
admission, it made ``no concessions to wealth, and it won't sacrifice
merit for diversity's sake'' \cite[p.~278]{golden}.  David Baltimore,
past president of Caltech and a member of the CNR, told Golden,
``People should be judged not by their parentage and wealth but by
their skills and ability,\,\ldots\ Any school that I'm associated
with, I want to be a meritocracy.'' Golden wrote that ``he assured me
that Caltech would never compromise its
standards''\cite[p.~284]{golden}.

Never say never. The era of uncompromising standards at Caltech has
come to an end.  The \emph{Los Angeles Times} reported on August 31,
2023 that Caltech is making historic changes to its admission
standards. ``In a groundbreaking step, the campus announced Thursday
that it will drop admission requirements for calculus, physics, and
chemistry courses for students who don't have access to them and offer
alternative paths.\,\ldots'' ``Data\ \ldots\ showed a significant
racial gap in access to those classes.''  Caltech's executive director
of undergraduate admissions explained the new policy in these terms
``\relax`I think that we're really in a time where institutions have
to decide if everything that they've been saying about diversity and
inclusion is true,' she said, noting that the challenge is especially
acute now that the U.S. Supreme Court has banned affirmative
action. `Is this something fundamental about who we are as an
institution\ \ldots\ or is this something that was just really nice
window dressing'\relax'' \cite{calculus}.

The action against Millikan has been one campaign within a much larger
political movement.  Millikan himself had this to say about those who
engage in mean-spirited attacks against America's
finest~\cite[70:535]{Millikan}:
\begin{quote}
To attempt to spread poison over the United States with respect to the
characters and motives of the finest, ablest and most public spirited
men whom America has recently produced is resorting to a method
which, it seems to me, all men of honesty and refinement can only
abhor and detest. -- Millikan
\end{quote}

The following remedies are recommended.  President Rosenbaum and the
Caltech Board of Trustees should rescind their endorsement of the CNR
report.  The report itself should be retracted for failing to meet the
minimal standards of accuracy and scholarship that are expected of
official documents issued by one of the world's great scientific
institutions.  Caltech should restore Robert Andrews Millikan to a
place of honor.

To be sure, Caltech has stirred up a hornets' nest.

\subsubsection*{Acknowledgments}
The author gives special
thanks to B. Charlesworth and other members of the Fisher Memorial
Trust, as well as J. Goodstein, P. Collopy, D. Kevles, M. Johnston, B. Palais,
R. Warne,  and the Caltech archives.  The author bears sole responsibility
for the article's content.

\bibliographystyle{plainnat}
\bibliography{all}

\end{document}